\documentclass[twocolumn]{revtex4} 

\usepackage{graphicx} 
\usepackage{pifont}
\usepackage{natbib}
\DeclareMathAlphabet{\mathpzc}{OT1}{pzc}{m}{it}

\newcommand {\lsim}{\mbox{$\:\stackrel{<}{_{\sim}}\:$} }
\newcommand {\gsim}{\mbox{$\:\stackrel{>}{_{\sim}}\:$} }

\def\be{\begin{equation}}
\def\ee{\end{equation}}
\def\bea{\begin{eqnarray}}
\def\eea{\end{eqnarray}}

\def\lcdm{$\Lambda$CDM}

\def\rhom{\rho_{m}}

\def\sol{\phantom{}_{\odot}}

\begin{document}
\title{Dark-Energy Dynamics Required to Solve the Cosmic Coincidence}

\author{Chas A. Egan}
\affiliation{Department of Astrophysics, School of Physics, 
University of New South Wales, Sydney, Australia}
\altaffiliation[Visiting the ]{Research School of Astronomy and Astrophysics, 
Australian National University}
\email{chas@mso.anu.edu.au}

\author{Charles H. Lineweaver}
\affiliation{Research School of Astronomy and Astrophysics,
Australian National University, Canberra, Australia}
\altaffiliation{Planetary Science Institute,
Australian National University}
\altaffiliation{Research School of Earth Sciences,
Australian National University}


\begin{abstract}
Dynamic dark energy (DDE) models are often designed to solve the cosmic coincidence 
(why, just now, is the dark energy density $\rho_{de}$, the same order of magnitude as the 
matter density $\rho_m$?) by guaranteeing $\rho_{de} \sim \rho_m$ for significant fractions 
of the age of the universe. 
However, such behaviour is neither sufficient nor necessary to solve the coincidence problem. 
Cosmological processes constrain the epochs during which observers can exist.
Therefore, what must be shown is that a significant fraction of observers see $\rho_{de} \sim \rho_m$. 
Precisely when, and for how long, must a DDE model 
have $\rho_{de} \sim \rho_{m}$ in order to solve the coincidence? 
We explore the coincidence problem in dynamic dark energy models using the temporal distribution of 
terrestrial-planet-bound observers. We find that 
any realistic DDE model 
which can be parameterized as $w=w_0+w_a(1-a)$ over a few e-folds, 
has $\rho_{de} \sim \rho_m$ for a significant fraction of 
observers
in the universe. This 
demotivates DDE models specifically designed to solve the coincidence using long or repeated 
periods of $\rho_{de} \sim \rho_m$.
\end{abstract}

\maketitle


\section{Introduction}

In 1998, using supernovae Ia as standard candles, \citet{Riess1998} and 
\citet{Perlmutter1999} revealed a recent and continuing epoch of cosmic acceleration - strong 
evidence that Einstein's cosmological constant $\Lambda$, or something else with comparable negative 
pressure $p_{de} \sim -\rho_{de}$, currently dominates the energy density of the universe 
\citep{Lineweaver1998}. $\Lambda$ is usually interpreted as the energy of zero-point quantum 
fluctuations in the vacuum \citep{ZelDovich1967,Durrer2007} with a constant equation of state 
$w \equiv p_{de}/\rho_{de} = -1$.
This necessary additional energy component, construed as $\Lambda$ or otherwise, has become 
generically known as ``dark energy'' (DE).

A plethora of observations have been used to constrain the free parameters of the
new standard cosmological model, \lcdm, in which $\Lambda$ \emph{does} play the role of the dark energy.
Hinshaw et al.\ \cite{Hinshaw2006} find that the universe is expanding at a rate of $H_0 = 71 \pm 4\ km/s/Mpc$; 
that it is spatially flat and therefore critically dense 
($\Omega_{tot 0} = \frac{\rho_{tot 0}}{\rho_{crit0}} = \frac{8 \pi G}{3 H_0^2} \rho_{tot 0} = 1.01 \pm 0.01$); 
and that the total density is comprised of contributions from 
vacuum energy ($\Omega_{\Lambda 0} = 0.74 \pm 0.02$), 
cold dark matter (CDM; $\Omega_{CDM 0} = 0.22 \pm 0.02$),
baryonic matter ($\Omega_{b 0} = 0.044 \pm 0.003$) and
radiation ($\Omega_{r 0} = 4.5 \pm 0.2 \times 10^{-5}$).
Henceforth we will assume that the universe is flat ($\Omega_{tot 0} = 1$) as predicted 
by inflation and supported by observations.

Two problems have been influential in moulding ideas about dark energy,
specifically in driving interest in alternatives to \lcdm.
The first of these problems is concerned with the smallness of the dark energy
density \citep{ZelDovich1967,Weinberg1989,Cohn1998}.  
Despite representing more than $70\%$ of the total energy of the universe, the 
current dark energy density is $\sim 120$ orders of magnitude smaller than 
energy scales at the end of inflation (or $\sim 80$ orders of magnitude smaller than 
energy scales at the end of inflation if this occurred at the GUT rather than Planck 
scale) \citep{Weinberg1989}. Dark energy 
candidates are thus challenged to explain why the observed DE density is so small. 
The standard idea, that the dark energy is the energy of zero-point quantum 
fluctuations in the true vacuum, seems to offer no solution to this problem.

The second cosmological constant problem \cite{Weinberg2000b,Carroll2001b,Steinhardt2003} 
is concerned with the near coincidence between the current cosmological matter density 
($\rho_{m0} \approx 0.26 \times \rho_{crit0}$) and the dark energy density 
($\rho_{de0} \approx 0.74 \times \rho_{crit0}$). In the standard \lcdm\ model, the 
cosmological window during which these components have comparable density is 
short (just $1.5$ e-folds of the cosmological scalefactor $a$)
since matter density dilutes as $\rho_{m} \propto a^{-3}$ while vacuum density 
$\rho_{de}$ is constant \citep{Lineweaver2007}.
Thus, even if one explains why the DE density is much less than the Planck density (the smallness 
problem) one must explain why we happen to live during the time when $\rho_{de} \sim \rho_m$.

The likelihood of this coincidence depends on the range of times during which we suppose we might 
have lived.
In works addressing the smallness problem, \citet{Weinberg1987,Weinberg1989,Weinberg2000} 
considered a multiverse consisting of a large number of big bangs, each with a different value 
of $\rho_{de}$. There he asked, suppose that we could have arisen in any one of these universes; 
What value of $\rho_{de}$ should we expect our universe to have? 
While Weinberg supposed we could have arisen in another universe, we are simply supposing that 
we could have arisen in another time
\emph{in our universe}. 
We ask, what time $t_{obs}$, and corresponding densities 
$\rho_{de}(t_{obs})$ and $\rho_{m}(t_{obs})$ should we expect to observe?
Weinberg's key realization was that not every universe was equally probable: those with smaller 
$\rho_{de}$ contain more Milky-Way-like galaxies and are therefore more hospitable 
\citep{Weinberg1987,Weinberg1989}. Subsequently, he, and other authors used the relative 
number of Milky-Way-like galaxies to estimate the distribution of observers as a function of 
$\rho_{de}$, and determined that our value of $\rho_{de}$ was indeed likely 
\citep{Efstathiou1995,Martel1998,Pogosian2007}. 
Our value of $\rho_{de}$ could have been found to 
be unlikely and this would have ruled out the type of multiverse being considered. 
Here we apply analogous reasoning to the cosmic coincidence problem. Our observerhood 
could not have happened at any time with equal probability \citep{Lineweaver2007}. By 
estimating the temporal distribution of observers we can determine whether the observation of 
$\rho_{de} \sim \rho_{m}$ was likely. If we find $\rho_{de} \sim \rho_{m}$ to be unlikely while 
considering a particular DE model, that will enable us to rule out that DE model.

In a previous paper \citep{Lineweaver2007}, we tested \lcdm\ in this way and found that 
$\rho_{de} \sim \rho_{m}$ is expected. In the present paper we apply this test to dynamic dark 
energy models to see what dynamics is required to solve the coincidence problem when the temporal
distribution of observers is being considered.


The smallness of the dark energy density has been anthropically explained in multiverse models
with the argument that in universes with much larger DE components, DE driven acceleration starts 
earlier and precludes the formation of galaxies and large scale structure. Such universes are 
probably devoid of observers \citep{Weinberg1987,Martel1998,Pogosian2007}.
A solution to the coincidence problem in this scenario was outlined by \citet{Garriga1999} who showed 
that if $\rho_{de}$ is low enough to allow galaxies to form, then observers in those galaxies will 
observe $r \sim 1$.

\begin{figure}[!tb]
       \begin{center}
               \includegraphics[scale=0.97]{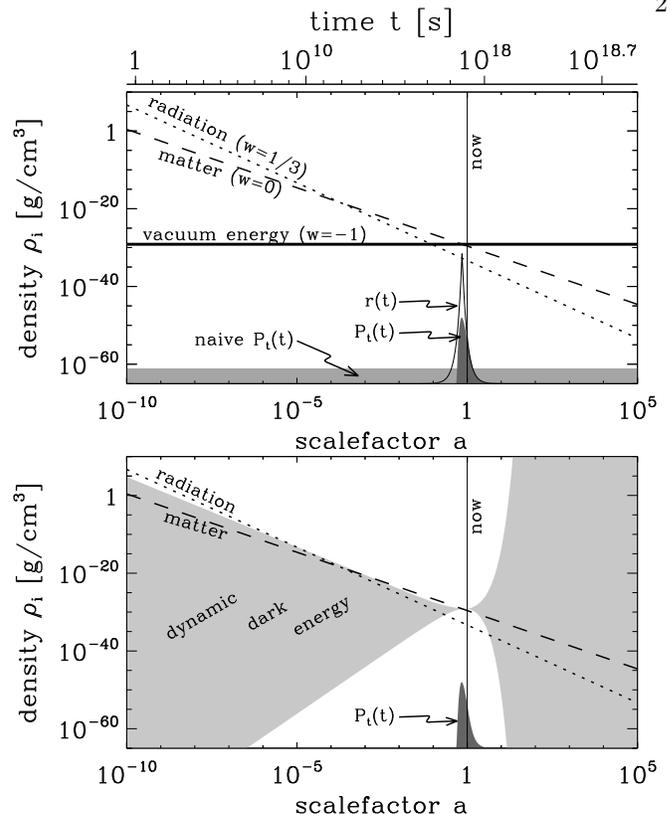}
               \caption{{\bf(Top)} The history of the energy density of the 
               universe according to standard \lcdm. The dotted line shows the
               energy density in radiation (photons, neutrinos and other relativistic
               modes). The radiation density dilutes as $a^{-4}$ as the universe
               expands. The dashed line shows the density in ordinary non-relativistic
               matter, which dilutes as $a^{-3}$. The thick solid line shows the energy 
               of the vacuum (the cosmological constant) which has remained constant since 
               the end of inflation. The thin solid peaked curve shows the proximity $r$ of 
               the matter density to the vacuum energy density (see Eq. \ref{eq:definer}).
               The proximity $r$ is only $\sim 1$ for a brief period in the $log(a)$ 
               history of the cosmos. Whether or not there is a coincidence problem
               depends on the distribution $P_t(t)$ for $t_{obs}$. If one naively
               assumes that we could have observed any epoch with equal probability (the 
               light grey shade) then we should not expect to observe $r$ as large as
               we do. If, however, $P_t(t)$ is based on an estimate of the temporal 
               distribution of observers (the dark grey shade) then $r_0 \approx 0.35$ is not 
               surprising, and the coincidence problem is solved under \lcdm\ \citep{Lineweaver2007}. 
               {\bf(Bottom)} The dark energy density history is modified in DDE models.
               Observational constraints on the dark energy density history are 
               represented by the light grey shade (details in Section \ref{constraints}).}
               \label{fig:illustrated_coincidence}
       \end{center}
\end{figure}

To quantify the time-dependent proximity of $\rho_m$ and $\rho_{de}$, we define a 
proximity parameter,
	\begin{equation} \label{eq:definer}
		r \equiv \min \left[ \frac{\rho_{de}}{\rho_m}, \frac{\rho_m}{\rho_{de}} \right],
	\end{equation}
which ranges from $r \approx 0$, when many orders of magnitude separate the two densities, 
to $r = 1$, when the two densities are equal. 
The presently observed value of this parameter is $r_0 = \frac{\rho_{m 0}}{\rho_{de 0}} \approx 0.35$. 
In terms of $r$, the coincidence problem is as follows. If we naively presume that the time of our 
observation $t_{obs}$ has been drawn from a distribution of times $P_t(t)$ spanning many decades of 
cosmic scalefactor, we find that the expected proximity parameter is $r \approx 0 \ll 0.35$. 
In the top panel of Fig.\ \ref{fig:illustrated_coincidence} we use a naive distribution for 
$t_{obs}$ that is constant in $\log(a)$ to illustrate how observing $r$ as large as $r_0 \approx 0.35$ 
seems unexpected.

In \citet{Lineweaver2007} we showed how the apparent severity 
of the coincidence problem strongly depends upon the distribution $P_t(t)$ from which $t_{obs}$ is 
hypothesized to have been drawn. Naive priors for $t_{obs}$, such as the one illustrated 
in the top panel of Fig.\ \ref{fig:illustrated_coincidence}, lead to naive conclusions.
Following the reasoning of \citet{Weinberg1987,Weinberg1989,Weinberg2000} we interpret $P_t(t)$ as 
the temporal distribution of observers. 
The temporal and 
spatial distribution of observers has been estimated using large ($10^{11} M_{\sol}$) galaxies 
\citep{Weinberg1987,Efstathiou1995,Martel1998,Garriga1999} and terrestrial planets 
\citep{Lineweaver2007} as tracers. The top panel of Fig.\ \ref{fig:illustrated_coincidence} shows 
the temporal distribution of observers $P_t(t)$ from \citet{Lineweaver2007}.

A possible extension of the concordance cosmological model that may explain the observed 
smallness of $\rho_{de}$ is the generalization of dark energy candidates to include dynamic 
dark energy (DDE) models such 
as quintessence, phantom dark energy, k-essence and Chaplygin gas. In these models the dark 
energy is treated as a new matter field which is approximately homogenous, and evolves as the 
universe expands. DDE evolution offers a mechanism for the decay of $\rho_{de}(t)$ from the expected
Planck scales ($10^{93}$ g/cm$^3$) in the early universe ($10^{-44}$ s) to the small value we 
observe today ($10^{-30}$ g/cm$^3$). The light 
grey shade in the bottom panel of Fig.\ \ref{fig:illustrated_coincidence} represents contemporary 
observational constraints on the DDE density history.
Many DDE models are designed to solve the coincidence problem 
by having $\rho_{de}(t) \sim \rho_{m}(t)$ for a large fraction of the history/future of the universe
\citep{Amendola2000,Dodelson2000,Sahni2000,Chimento2000,Zimdahl2001,Sahni2002,Chimento2003,Ahmed2004,
Franca2004,Mbonye2004,delCampo2005,Guo2005,Olivares2005,Pavon2005,Scherrer2005,Zhang2005,delCampo2006,
Franca2006,Feng2006,Nojiri2006,Amendola2006,Amendola2007,Olivares2007,Sassi2007}.
With $\rho_{de} \sim \rho_{m}$ for extended or repeated periods the hope is to ensure that $r \sim 1$ 
is expected.

Our main goal in this paper is to take into account the temporal distribution of observers to 
determine when, and for how long, a DDE model must have $\rho_{de} \sim \rho_{m}$ in order to solve 
the coincidence problem? Specifically, we extend the work of \citet{Lineweaver2007} to find out for
which cosmologies (in addition to \lcdm) the coincidence problem is solved when the temporal
distribution of observers is considered.
In doing this we answer the question, Does a dark energy model fitting contemporary 
constraints on the density $\rho_{de}$ and the equation of state parameters, necessarily solve 
the cosmic coincidence?
Both positive and negative answers have interesting consequences. An answer in the affirmative 
will simplify considerations that go into DDE modeling: any DDE model in agreement with current cosmological 
constraints has $\rho_{de} \sim \rho_m$ for a significant fraction of observers.
An answer in the negative would yield a new 
opportunity to constrain the DE equation of state parameters \emph{more strongly} than contemporary 
cosmological surveys. 

A different coincidence problem arises when the time of observation is conditioned on 
and the parameters of a model are allowed to slide. The tuning of parameters and 
the necessity to include ad-hoc physics are large problems for many current dark energy models.
This paper does not address such issues, and the interested reader is referred to \citet{Hebecker2001},
\citet{Bludman2004} and \citet{Linder2006}. In the coincidence problem addressed here we let the time 
of observation vary to see if $r(t_{obs}) \ge 0.35$ is unlikely according to the model.

In Section \ref{solving} we present several examples of DDE models
used to solve the coincidence problem. An overview of observational constraints on 
DDE is given in Section \ref{constraints}. 
In Section \ref{observers} we estimate the temporal distribution of observers. 
Our main analysis is presented in Section \ref{results}. Our main result - that the coincidence
problem is solved for all DDE models fitting observational constraints - is
illustrated in Fig.\ \ref{fig:severities}. Finally, in Section \ref{conclusion}, we end with a 
discussion of our results, their implications and potential caveats. 


\section{Dynamic Dark Energy Models in the Face of the Cosmic Coincidence} \label{solving}

\begin{figure*}[!tb]
       \begin{center}
               \includegraphics[scale=0.8]{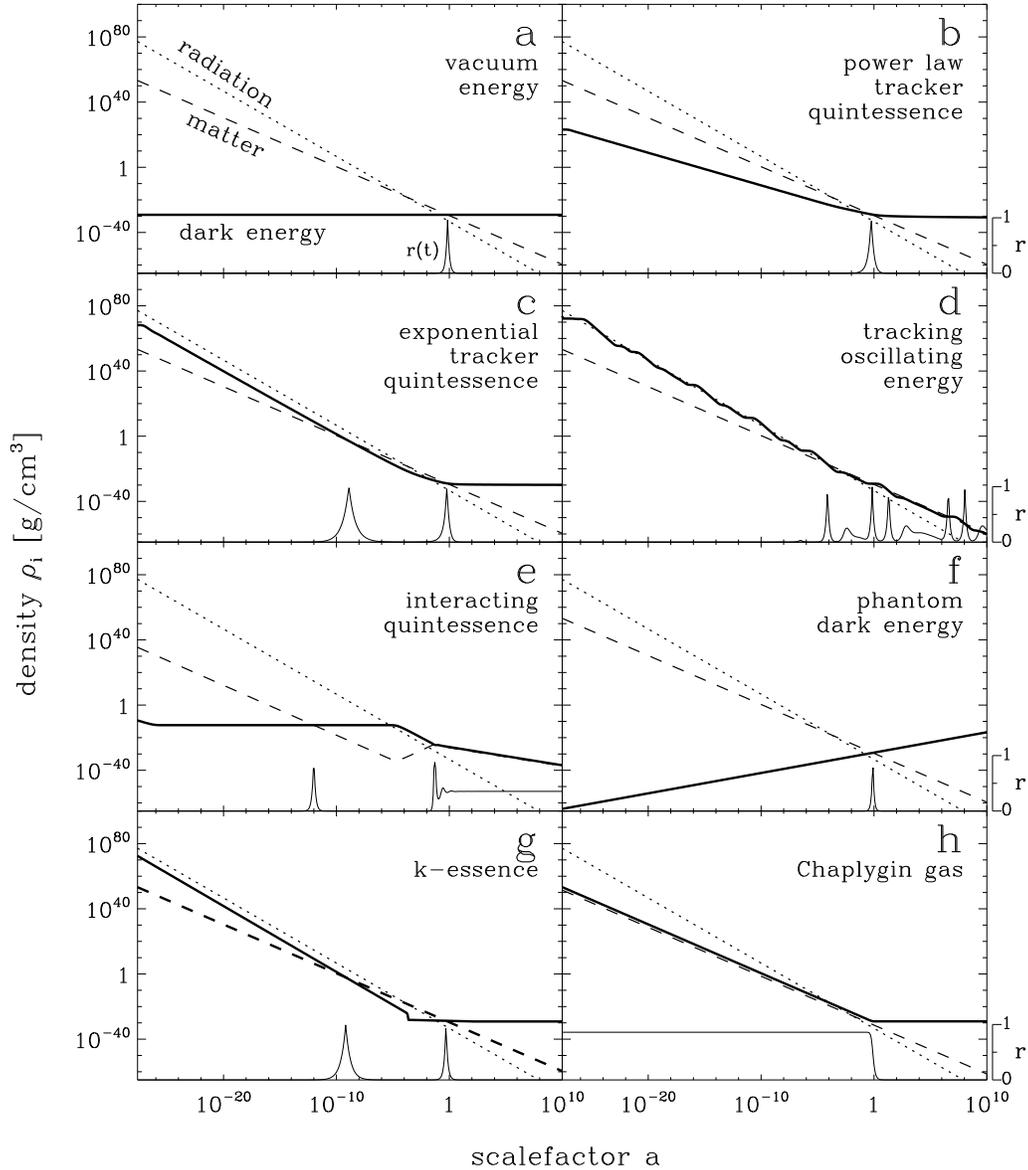}
               \caption{The energy density history of the universe according to \lcdm\ (panel a), and seven
				DDE models selected from the literature (see text for references). In each panel the
				radiation and matter densities are the dotted and dashed lines respectively. The DE
				density is given by the thick black line. The
				proximity parameter $r$ is given by the thin black line at the base of each panel.
				Of the DDE models shown here, tracker quintessence and k-essence (panels b, c and g) have $r \sim 1$
				for a small fraction of the life of the universe (whether the abscissa is $t$, $\log(t)$, $a$,
				$\log(a)$, or any other of a large number of measures).
				On the other hand, tracking oscillating energy, interacting quintessence, phantom DE 
				and Chaplygin gas (panels d, e, f and h) exhibit $r \sim 1$
				for a large fraction of the life of the universe. 
				For the phantom DE example (panel f) this is true 
				in $t$, but not 
				in $a$ or $\log(a)$.
				In phantom models the future universe grows super-exponentially to $a=\infty$ (a ``big-rip'') 
				shortly after matter-DE equality. Thus the universe spends a large fraction of \emph{time} 
				with $r \sim 1$, however this is is not seen in $\log(a)$-space. 
				For each of the models in this figure, numerical values for free parameters were chosen
				to crudely fit observational constraints and are given in Appendix \ref{paramvals}.}
               \label{fig:quintessence_energy}
       \end{center}
\end{figure*}

Though it is beyond the scope of this article to provide a complete review of DDE (see 
\citet{Copeland2006,Szydlowski2006}), here we give a few representative examples in 
order to set the context and motivation of our work.
Fig.\ \ref{fig:quintessence_energy} illustrates density histories typical of tracker quintessence, 
tracking oscillating energy, interacting quintessence, phantom dark energy, k-essence, and Chaplygin 
gas. They are discussed in turn below.

\subsection{Quintessence}

In quintessence models the dark energy is interpreted as a homogenous scalar field with 
Lagrangian density ${\mathcal L}(\phi,X) = \frac{1}{2} \dot{\phi}^2 - V(\phi)$
\citep{Ozer1987,Ratra1988,Ferreira1998,Caldwell1998,Steinhardt1999,Zlatev1999,Dalal2001}.
The evolution of the quintessence field and of the cosmos depends on the postulated
potential $V(\phi)$ of the field and on any postulated interactions. In general, quintessence 
has a time-varying equation of state 
$w = \frac{p_{de}}{\rho_{de}} = \frac{\dot{\phi}^2 / 2 - V(\phi)}{\dot{\phi}^2 / 2 + V(\phi)}$.
Since the kinetic term $\dot{\phi}^2 / 2$ cannot be negative, the equation of state is restricted 
to values $w \ge -1$. Moreover, if the potential $V(\phi)$ is non-negative then $w$ is also 
restricted to values $w \le +1$.

If the quintessence field only interacts gravitationally then energy density evolves as 
$\frac{\delta \rho_{de}}{\rho_{de}} = -3(w+1) \frac{\delta a}{a}$ and the restrictions 
$-1 \le w \le +1$ mean $\rho_{de}$ decays (but never faster than $a^{-6}$) or remains constant 
(but never increases). 

\subsubsection{Tracker Quintessence}

Particular choices for $V(\phi)$ lead to interesting attractor solutions which can be exploited 
to make $\rho_{de}$ scale (``track'') sub-dominantly with $\rho_r + \rho_m$.

The DE can be forced to transit to a $\Lambda$-like ($w \approx -1$) state at any time by 
fine-tuning $V(\phi)$. In the $\Lambda$-like state $\rho_{de}$ overtakes $\rho_m$ and dominates 
the recent and future energy density of the universe. We illustrate tracker quintessence in Fig.\ 
\ref{fig:quintessence_energy} using a power law potential $V(\phi) = M \phi^{-\alpha}$ (panel b)
\citep{Ratra1988,Caldwell1998,Zlatev1999} and an exponential potential $V(\phi) = M \exp(1/\phi)$ 
(panel c) \citep{Dodelson2000}.

The tracker paths are attractor solutions of the equations governing the evolution of the field. 
If the tracker quintessence field is initially endowed with a density off the tracker path (e.g.\ 
an equipartition of the energy available at reheating) its density quickly approaches and joins 
the tracker solution.

\subsubsection{Oscillating Dark Energy}
\citet{Dodelson2000} explored a quintessence potential with oscillatory perturbations 
$V(\phi) = M \exp(- \lambda \phi) \left[1 + A \sin(\nu \phi)\right]$. They refer to
models of this type as tracking oscillating energy. Without the
perturbations (setting $A=0$) this potential causes exact tracker behaviour: the 
quintessence energy decays as $\rho_r + \rho_m$ and never dominates. With the perturbations
the quintessence energy density oscillates about $\rho_r + \rho_m$ as it decays (Fig.\ 
\ref{fig:quintessence_energy}d). The quintessence energy 
dominates on multiple occasions and its equation of state varies continuously between 
positive and negative values. One of the main motivations for tracking oscillating energy 
is to solve the coincidence problem by ensuring that $\rho_{de} \sim \rho_{m}$ or $\rho_{de} \sim \rho_{r}$
at many times in the past or future.

It has yet to be seen how such a potential might
arise from particle physics. Phenomenologically similar cosmologies have been discussed 
in \citet{Ahmed2004,Yang2005,Feng2006}.

\subsubsection{Interacting Quintessence}
Non-gravitational interactions between the quintessence field and matter fields might allow energy 
to transfer between these components. Such interactions are not forbidden by any known symmetry
\cite{Amendola2000b}. The primary motivation for the exploration of interacting dark energy models 
is to solve the coincidence problem. In these models the present matter/dark energy density 
proximity $r$ may be constant \citep{Amendola2000,Zimdahl2001,Amendola2003,Franca2004,Guo2005,
Olivares2005,Pavon2005,Zhang2005,Franca2006,Amendola2006,Amendola2007,Olivares2007} or slowly 
varying \citep{delCampo2005,delCampo2006}. 
 
We plot a density history of the interacting quintessence model of \citet{Amendola2000} in Fig.\ 
\ref{fig:quintessence_energy}e. This model is characterized by a DE potential $V(\phi)=A \exp[B \phi]$ 
and DE-matter interaction term $Q = -C \rho_m \dot{\phi}$, specifying the rate at which energy 
is transferred to the matter fields. The free parameters were tuned such that radiation domination 
ends at $a=10^{-5}$ and that $r_{t \rightarrow \infty} = 0.35$.

\subsection{Phantom Dark Energy}
The analyses of \citet{Riess2004} and \citet{Wood-Vasey2007} have mildly ($\sim 1 \sigma$) 
favored a dark energy equation of state $w_{de} < -1$. 
These values are unattainable by standard quintessence models
but can occur in phantom dark energy models \citep{Caldwell2002}, in which kinetic energies
are negative. The energy density in the phantom field \emph{increases} with 
scalefactor, typically leading to a future ``big rip'' singularity where the scalefactor 
becomes infinite in finite time. Fig.\ \ref{fig:quintessence_energy}f shows the density 
history of a simple phantom model with a constant equation of state $w=-1.25$. The big rip
($a=\infty$ at $t=57.5$ Gyrs) is not seen in $\log(a)$-space.

\citet{Caldwell2003} and \citet{Scherrer2005} have explored how phantom models may solve 
the coincidence problem: since the big rip is triggered by the onset of DE domination, 
such cosmologies spend a significant fraction of their total time with $r$ large. 
For the phantom model with $w=-1.25$ (Fig.\ \ref{fig:quintessence_energy}f) \citet{Scherrer2005}
finds $r > 0.1$ for $12\%$ of the total lifetime of the universe. Whether this solves the 
coincidence or not depends upon the prior probability distribution $P_t(t)$ for the time of 
observation. \citet{Caldwell2003} and \citet{Scherrer2005} implicitly assume that the temporal 
distribution of observers is constant in time (i.e.\ $P_t(t)=\textrm{constant}$). For this prior 
the coincidence problem \emph{is} solved because the chance of observing $r \ge 0.1$ is large 
($12\%$). Note that for the ``naive $P_t(t)$'' prior shown in Fig.\ \ref{fig:illustrated_coincidence}, 
the solution of \citet{Caldwell2003} and \citet{Scherrer2005} fails because $r > 0.1$ is brief in 
$\log(a)$-space. It fails in this way for many other choices of $P_t(t)$ including, for example, 
distributions constant in $a$ or $\log(t)$.

\subsection{K-Essence}
In k-essence the DE is modeled as a scalar field with non-canonical kinetic energy 
\citep{Chiba2000,Armendariz-Picon2000,Armendariz-Picon2001,Malquarti2003}. Non-canonical
kinetic terms can arise in the effective action of fields in string and supergravity 
theories. Fig.\ \ref{fig:quintessence_energy}g shows a density history typical of 
k-essence models. This particular model is from \citet{Armendariz-Picon2001} and \citet{Steinhardt2003}. 
During radiation domination the k-essence field tracks radiation sub-dominantly (with 
$w_{de}=w_r=1/3$) as do some of the other models in Fig.\ \ref{fig:quintessence_energy}. 
However, 
no stable tracker solution exists for $w_{de}=w_m (=0)$. Thus after radiation-matter equality, 
the field is unable to continue tracking the dominant component, and is driven to another 
attractor solution (which is generically $\Lambda$-like with $w_{de} \approx -1$). 
The onset of DE domination was recent in k-essence models because matter-radiation equality 
prompts the transition to a $\Lambda$-like state. K-essence thereby 
avoids fine-tuning in any particular numerical parameters, but the Lagrangian 
has been constructed ad-hoc.

\subsection{Chaplygin Gas}
A special fluid known as Chaplygin gas motivated by braneworld cosmology may be able to 
play the role of dark matter \emph{and} the dark energy \citep{Bento2002,Kamenshchik2001}. 
Generalized Chaplygin gas has the equation of state $p_{de}=-A \rho_{de}^{-\alpha}$ 
which behaves like pressureless dark matter at early times ($w_{de} \approx 0$ when $\rho_{de}$ 
is large), and like vacuum energy at late times ($w_{de} \approx -1$ when $\rho_{de}$ is small). 
In Fig.\ \ref{fig:quintessence_energy}h we show an example with 
$\alpha=1$.

\subsection{Summary of DDE Models}
Two broad classes of DDE models emerge from our comparison:
\begin{enumerate}
\item In \lcdm, tracker quintessence and k-essence models, 
the dark energy density is vastly different from the matter density
for most of the lifetime of the universe (panels a, b, c, g of Fig.\ \ref{fig:quintessence_energy}). 
The coincidence problem can only be solved if the probability distribution $P_t(t)$ for the time of 
observation is narrow, and overlaps significantly with an $r \sim 1$ peak. If $P_t(t)$ 
is wide, e.g.\ constant over the life of the universe in $t$ or $\log(t)$, then
observing $r \sim 1$ would be unlikely in these models and the coincidence problem \emph{is not} 
resolved.
\item Tracking oscillating energy, interacting quintessence, phantom models 
and Chaplygin gas models (panels d, e, f, h of Fig.\ \ref{fig:quintessence_energy}) employ 
mechanisms to ensure that $r \sim 1$ for large fractions of the life of the universe. In these 
models the coincidence problem may be solved for a wider range of $P_t(t)$ including, depending
on the DE model, distributions that are constant over the whole life of the universe 
in $t$, $\log(t)$, $a$ or $\log(a)$.
\end{enumerate}
The importance of an estimate of the distribution $P_t(t)$ is highlighted: such an 
estimate will either rule out models of the first category because they do not solve the coincidence
problem, or demotivate models of the second because their mechanisms are unnecessary to solve the
coincidence problem. This analysis does not addressed the problems associated with fine-tuning, 
initial conditions or ad hoc mechanisms of many DDE models \citep{Hebecker2001,Bludman2004,Linder2006}.
 
We leave this line of enquiry temporarily to discuss contemporary \emph{observational} constraints on the dark 
energy density history, because we wish to test what DE dynamics are required to solve the
coincidence, beyond those which models must exhibit to satisfy standard cosmological observations.


\section{Current Observational Constraints on Dynamic Dark Energy} \label{constraints}

\begin{figure*}[!hbt]
       \begin{center}
               \includegraphics[scale=0.8]{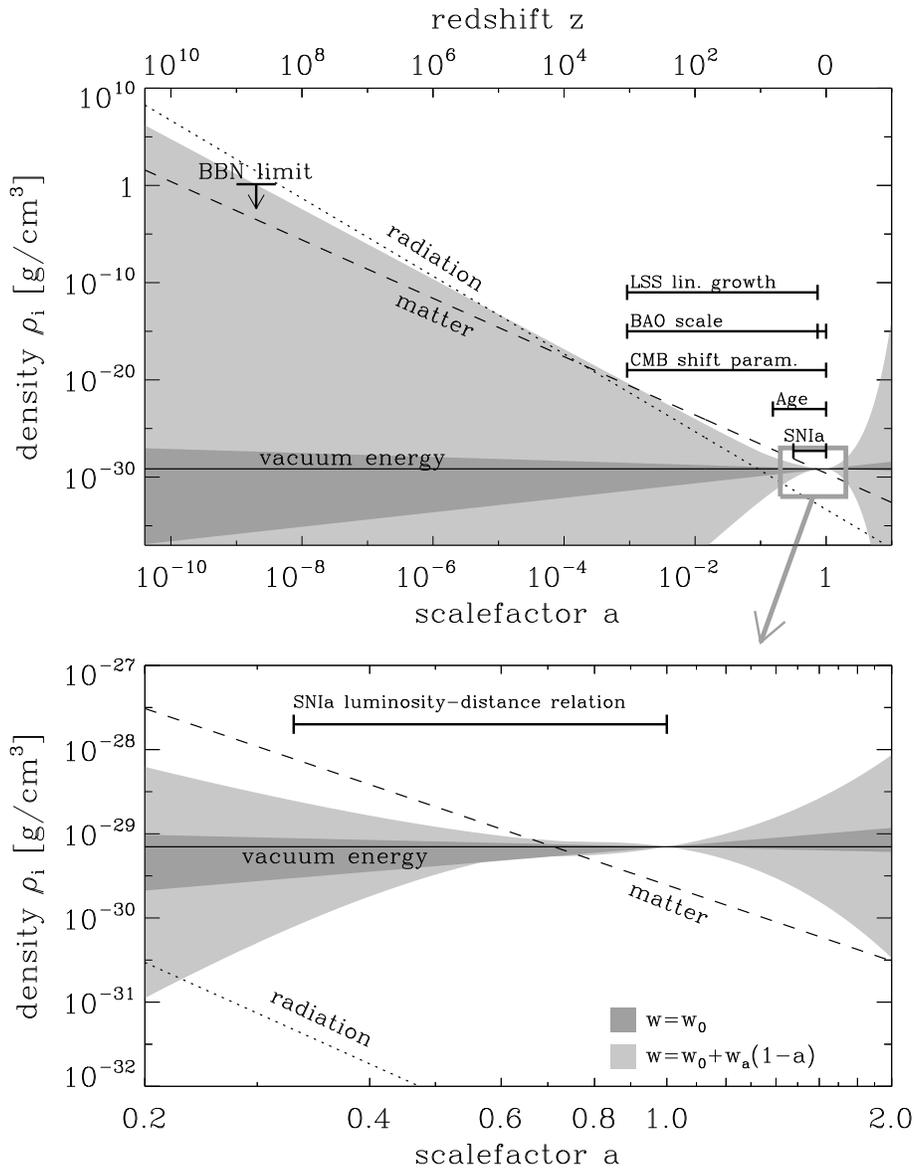}
               \caption{The energy densities of radiation $\rho_r$, matter $\rho_m$ and 
               	the cosmological constant $\rho_{\Lambda}$ are shown as a function of
	           scalefactor, by the dotted, dashed, and solid lines respectively. 
	           Cosmological probes of dark energy include SNIa, CMB, BAO, the LSS linear
	           growth factor and constraints from BBN (see text). Each of these probes is sensitive 
	           to the effects of dark energy over different redshift intervals, as indicated.
	           The light grey band envelopes $w_0$-$w_a$-parameterized DDE models allowed at 
	           $<2 \sigma$ by \citet{Davis2007} (the contour in $w_0-w_a$ space is shown 
	           explicitly in Fig.\ \ref{fig:severities}). The dark grey band envelopes 
	           $w_0$-parameterized DDE models ($w_a = 0$ assumed) allowed at $<2 \sigma$ by 
	           \citet{Wood-Vasey2007}. The constraint is $w = -1.09 \pm 0.16$ at $2 \sigma$.
	           }
               \label{fig:SN1aConstraints}
       \end{center}
\end{figure*}

\subsection{Supernovae Ia}

Observationally, possible dark energy dynamics is explored almost solely using measurements of 
the cosmic expansion history.
Recent cosmic expansion is directly probed by using type Ia supernova 
(SNIa) as standard candles \citep{Riess1998,Perlmutter1999}. Each observed SNIa provides an
independent measurement of the luminosity distance $d_l$ to the redshift of the supernova $z_{SN}$.
The luminosity distance to $z_{SN}$ is given by
\be \label{eq:lumdist}
d_l(z_{SN}) = (1+z_{SN}) \frac{c}{H_0} \int_{z=0}^{z_{SN}} \frac{d z}{E(z)}
\ee 
where
\bea
E(z) & = & \frac{H(z)}{H_0} \label{eq:e} \\
     & = & \left[ \Omega_{r 0}(1+z)^4 + \Omega_{m 0}(1+z)^3 + \Omega_{de 0} 
     \frac{\rho_{de}(z)}{\rho_{de 0}}\right]^{\frac{1}{2}} \nonumber
\eea 
and thus depends on $H_0$, $\Omega_{m 0}$, and the evolution of the dark energy $\rho_{de}(z) / \rho_{de 0}$. 
The radiation term, irrelevant at low redshifts, can be dropped from Equation \ref{eq:e}. 
$\Omega_{de 0}$ is a dependent parameter due to flatness ($\Omega_{de 0} = 1-\Omega_{m 0}$).
Contemporary datasets include $\sim 200$ supernovae at redshifts $z_{SN} \le 2.16$ ($a \ge 0.316$) 
\citep{Riess2007,Wood-Vasey2007} and provide an effective continuum of constraints on the expansion 
history over that range \citep{Wang2005,Wang2006}. 
The redshift range probed by SNIa is indicated in both panels of Fig.\ \ref{fig:SN1aConstraints}. 

\subsection{Cosmic Microwave Background}

The first peak in the cosmic microwave background (CMB) temperature power spectrum corresponds 
to density fluctuations 
on the scale of the sound horizon at the time of recombination. Subsequent peaks correspond to 
higher-frequency harmonics. The locations of these peaks in $l$-space depend on the comoving 
scale of the sound horizon at recombination, and the angular distance to recombination.
This is summarized by the so-called CMB shift parameter $R$ \citep{Efstathiou1999,Elgaroy2007} 
which is related to the cosmology by
\be \label{eq:cmbshift}
 	R = \sqrt{\Omega_{m0}} \int_{z=0}^{z_{rec}} \frac{d z}{E(z)}
\ee
where $z_{rec} \approx 1089$ \citep{Spergel2006} is the redshift of recombination. 
The 3-year WMAP data gives a shift parameter $R=1.71 \pm 0.03$ \citep{Davis2007,Spergel2006}.
Since the dependence of Equation \ref{eq:cmbshift} on $H_0$ and $\Omega_{m 0}$ differs from that 
of Equation \ref{eq:lumdist}, measurements of the CMB shift parameter can be used to break 
degeneracies between $H_0$, $\Omega_{m 0}$ and DE evolution in the analysis of SNIa.
In the top panel of Fig.\ \ref{fig:SN1aConstraints} we represent the CMB observations using a 
bar from $z=0$ to $z_{rec}$.

\subsection{Baryonic Acoustic Oscillations and Large Scale Structure}

As they imprinted acoustic peaks in the CMB, the baryonic oscillations at recombination were
expected to leave signature wiggles - baryonic acoustic oscillations (BAO) - in the power spectrum 
of galaxies \citep{EisensteinHu1998}. These were detected with significant confidence in the 
SDSS luminous red galaxy power spectrum \citep{Eisenstein2005}. The expected BAO scale depends on the scale of 
the sound horizon at recombination, and on transverse and radial scales at the mean redshift $z_{BAO}$, of 
galaxies in the survey.
\citet{Eisenstein2005} measured the quantity 
\be
A(z_{BAO}) = \frac{\sqrt{\Omega_{m 0}}}{E(z_{BAO})^{\frac{1}{3}}} \left[ \frac{1}{z_{BAO}} \int_{z=0}^{z_{BAO}} \frac{dz}{E(z)} \right]^{\frac{2}{3}}
\ee
to have a value $A(z_{BAO}=0.35) = 0.469 \pm 0.017$,
thus constraining the matter density and the dark energy evolution parameters in a configuration
which is complomentary to the CMB shift parameter and the SNIa luminosity distance relation. 
Ongoing BAO projects have been designed specifically to produce stronger constraints on the 
dark energy equation of state parameter $w$. For example, WIGGLEZ \citep{Glazebrook2007} will use a 
sample of high-redshift galaxies to measure the BAO scale at $z_{BAO} \approx 0.75$. As well as 
reducing the effects of non-linear clustering, this redshift is at a larger angular distance, making 
the observed scale more sensitive to $w$.
Constraints from the BAO scale depend on the evolution of the universe from $z_{rec}$ to 
$z_{BAO}$ to set the physical scale of the oscillations. They also depend on the evolution of the
universe from $z_{BAO}$ to $z=0$, since the observed angular extent of the oscillations depends on 
this evolution. The bar representing BAO scale observations in the top panel of Fig.\ 
\ref{fig:SN1aConstraints} indicates both these regimes.

The amplitude of the BAOs - the amplitude of the large scale structure (LSS) power spectrum - is 
determined by the amplitude of the power spectrum at recombination, and how much those fluctuations 
have grown (the transfer function) between $z_{rec}$ and $z_{BAO}$. By comparing the recombination 
power spectrum (from CMB) with the galaxy power spectrum, the LSS linear growth factor can be measured 
and used to constrain the expansion history of the universe (independently of the BAO scale) over 
this redshift range. In practice, biases hinder precise normalization of the galaxy power spectrum, 
weakening this technique. The range over which this technique probes the DE is indicated in Fig.\ 
\ref{fig:SN1aConstraints}.

\subsection{Ages}

Cosmological parameters from SN1a, CMB, LSS, BAO and other probes allow us to calculate the current 
age of the universe to be $13.8 \pm 0.1$ \citep{Hinshaw2006} assuming \lcdm. Uncertainties on the age 
calculated in this way grow dramatically if we drop the assumption that the DE is vacuum energy ($w=-1$). 

An independent lower limit on the current age of the universe is found by estimating the ages of the 
oldest known globular clusters \citep{Hansen2004}. 
These observations rule out models which predict the universe to be younger than 
$12.7 \pm 0.7 \; \textrm{Gyrs}$ ($2 \sigma$ confidence):
\bea
t_0 & = & H_0^{-1} \int_{z=0}^{\infty} \frac{dz}{(1+z) E(z)} \\
 	& \gsim & 12.7 \pm 0.7 \; \textrm{Gyrs}. \nonumber
\eea
Other objects can also be used to set this age limit \citet{Lineweaver1999}, but generally less 
successfully due to uncertainties in dating techniques.

Assuming \lcdm, an age of $12.7$ Gyrs corresponds to a redshift of $z \approx 5.5$. Contemporary age 
measurements are sensitive to the dark energy content from $z \approx 5.5$ to $z=0$. In the top panel
of Fig.\ \ref{fig:SN1aConstraints} we show this redshift interval. The evolution and energy content 
of the universe before $12.7$ Gyrs ago is not probed by these age constraints.

\subsection{Nucleosynthesis}

In addition to the constraints on the expansion history (SN1a, CMB, BAO and $t_0$) we 
know that $\rho_{de}/\rho_{tot} < 0.045$ (at $2 \sigma$ confidence) during Big Bang Nucleosynthesis (BBN)
\citep{Bean2001b}. Larger dark energy densities imply a higher expansion rate at that epoch 
($z \sim 6 \times 10^{8}$) which would result in a lower neutron to proton ratio, conflicting with the 
measured helium abundance, $Y_{\textrm{He}}$.

\subsection{Dark Energy Parameterization}

Because of the variety of proposed dark energy models, it has become usual to summarize
observations by constraining a parameterized time-varying equation of state. Dark 
energy models are then confronted with observations in this parameter space.
The unique zeroth order parameterization of $w$ is $w=w_0$ (a constant), with $w=-1$ 
characterizing the cosmological constant model. 
The observational data can be used to constrain the first derivative of $w$. 
This additional dimension in the DE parameter space may be useful in distinguishing
models which have the same $w_0$. From an observational standpoint, the 
obvious choice of 1st order parameterization is $w(z)=w_0+\frac{dw}{dz}z$ \citep{diPietro2003}. 
This is rarely used today since currently considered DDE models are poorly portrayed by 
this functional form. The most popular parameterization is $w(a)=w_0+w_a(1-a)$ 
\cite{Albrecht2006,Linder2006b}, which does not diverge at high redshift. 

\citet{Linder2005} have argued that the extension of this approach to second order, e.g.\ 
$w(a)=w_0+w_a(1-a)+w_{aa}(1-a)^2$, is not motivated by current DDE models. Moreover, they 
have shown that next generation observations are unlikely to be able to distinguish the
quadratic from a linear expansion of $w$. \citet{Riess2007} have illustrated this recently 
using new SN1a. 

An alternative technique for exploring the history of dark energy is to constrain $w(z)$ or 
$\rho_{de}(z)$ in independent redshift bins. This technique makes fewer assumptions about 
the specific shape of $w(z)$. In the absence of any strongly motivated parameterization of 
$w(z)$ this bin-wise method serves as a good reminder of how little we actually know from
observation.
Using luminosity distance measurements from SNIa, DE evolution has been constrained in this way
in $\Delta z \sim 0.5$ bins out to redshift $z_{SN} \sim 2$ \citep{Wang2004b,Huterer2005,Riess2007}.
In the future, BAO measurements at various redshifts may contribute to these constraints, 
however $z_{BAO}$ will probably never be larger than $z_{SN}$.
Moreover, because the recombination redshift 
$z_{rec} \approx 1089$ is fixed, only the cumulative effect (from $z=z_{rec}$ to $z=0$) of the 
DE can be measured with the CMB and LSS linear growth factor. With only this single data point 
above $z_{SN}$, the bin-wise technique effectively degenerates to a parameterized analysis at 
$z > z_{SN}$.

\subsection{Summary of Current DDE Constraints}

If one assumes the popular $w_0 - w_a$ parameterization until last 
scattering, then all cosmological probes can be combined to constrain $w_0$ and 
$w_a$. In a recent analysis of SN1a, CMB and BAO observations, \citet{Davis2007} found 
$w_0 = -1.0 \pm 0.4$ and $w_a = -0.4 \pm 1.8$ at $2 \sigma$ confidence (the
contour is shown in Fig.\ \ref{fig:severities}). Using the same 
observations, \citet{Wood-Vasey2007} assumed $w_a = 0$ and found $w = w_0 = -1.09 \pm 0.16$
($2 \sigma$).

The evolution of $\rho_{de}$ is related to $w$ by covariant energy conservation 
\citep{Carroll2004book}
\be
\frac{\delta \rho_{de}}{\rho_{de}} = - 3 \left( w(a)+1 \right) \frac{\delta a}{a}.
\ee
The dark energy density corresponding to the $w_0 - w_a$ parameterization of $w$ is
thus given by 
\be
\rho_{de}(z) = \rho_{de 0}\ e^{3 w_a (a - 1)}\ a^{-3 (1 + w_0 + w_a)}.
\ee

The cosmic energy density history is illustrated in Fig.\ \ref{fig:SN1aConstraints}. Radiation and
matter densities steadily decline as the dotted and dashed lines. With the DE equation of 
state parameterized as $w(a)=w_0+w_a(1-a)$, its density history is constrained to the light-grey 
area \citep{Davis2007}. If the evolution of $w$ is negligible, i.e.\ we condition on $w_a \approx 0$,
then $w(a) \approx w_0$ and the DE density history 
lies within the dark-grey band \citep{Wood-Vasey2007}. If the dark energy is pure vacuum energy 
(or Einstein's cosmological constant) then $w=-1$ and its density history is given by the 
horizontal solid black line. 


\section{The Temporal Distribution of Observers} \label{observers}

The energy densities $\rho_r$, $\rho_m$ and $\rho_{de}$, and the proximity parameter $r$ 
we imagine we might 
have observed, depend on the distribution $P_t(t)$ from which
we imagine 
our time of observation $t_{obs}$ 
has been drawn. 
What we can expect to 
observe must be restricted by the conditions necessary for our presence as observers \citep{Carter1974}.
Thus, for example, it is meaningless to suppose we might have lived during inflation, or during radiation
domination, or before the first atoms \cite{Dicke1961}. 

We can, however, suppose that we are randomly selected cosmology-discovering observers, and we can expect 
our observations of $\rho_m$ and $\rho_{de}$ to be typical of observations made by such observers.
This is Vilenkin's principle of mediocrity \cite{Vilenkin1995a}. Accordingly, the distribution 
$P_t(t)$ for the time of observation $t_{obs}$ is proportional to the temporal distribution of 
cosmology-discovering observers (referred to henceforth as simply ``observers''). Thus to solve the coincidence 
problem one must show that the proximity parameter we measure, $r_0$, is typical of those measured by 
other observers.

The most abundant elements in the cosmos are hydrogen, helium, oxygen and carbon \citep{Pagel1997}. 
In the past decade $>200$ extra solar planets have been observed via doppler, transit or microlensing
methods. 
Extrapolation of current patterns in planet mass and orbital period are consistent with the idea that 
planetary systems like our own are common in the universe \cite{Lineweaver2003b}. 
All this does not necessarily imply that observers are 
common, but it does support the idea 
that terrestrial-planet-bound carbon-based 
observers, even if rare, may be the \emph{most common} observers. In the following estimation of $P_t(t)$ 
we consider only observers bound to terrestrial planets.

\subsection{First the Planets...} 

\begin{figure}[!hbt]
       \begin{center}
               \includegraphics[scale=0.52]{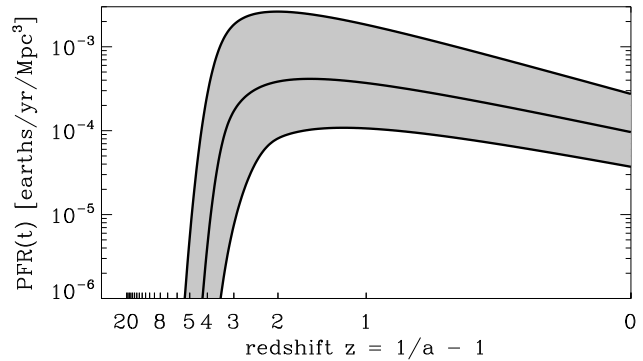}
               \caption{The terrestrial planet formation rate as estimated by \citet{Lineweaver2001}.
				It is based on a compilation of SFR measurements and has been corrected for 
				the low metallicity of the early universe, which prevents the terrestrial planet 
				formation rate from rising as quickly as the stellar formation rate at $z \gsim 4$.}
               \label{fig:tpfhz}
       \end{center}
\end{figure}

\citet{Lineweaver2001} estimated the terrestrial planet formation rate (PFR) by making 
a compilation of measurements of the cosmic star formation rate (SFR) and suppressing a 
fraction of the early stars $f(t)$ to correct for the fact that the metallicity was too low for 
those early stars to host terrestrial planetary systems,
	\begin{equation}
		PFR(t) = const \times SFR(t) \times f(t).
	\end{equation}
In Fig.\ \ref{fig:tpfhz} we plot the PFR reported by \citet{Lineweaver2001} as a function of 
redshift, $z=\frac{1}{a}-1$. As illustrated in the figure, there is large uncertainty in the normalization 
of the formation history. The number of stars orbited by terrestrial planets normalizes the distribution of 
observers but, importantly, does not shift the distribution in time. Thus our analysis will not depend on 
the normalization of this function and this 
uncertainty \emph{will not} propagate into our analysis. There are also uncertainties 
in the location of the turnover at high redshift, and in the slope of the formation history 
at low redshift - both of these \emph{will} affect our results.

\begin{figure}[!hbt]
       \begin{center}
               \includegraphics[scale=0.52]{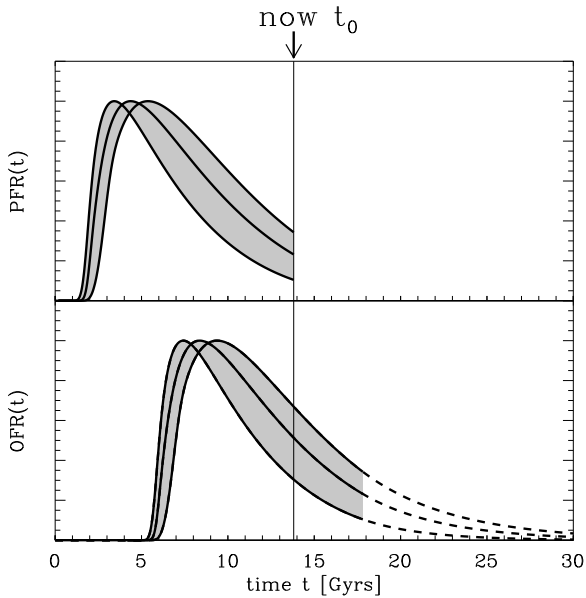}
               \caption{The terrestrial planet formation from Fig.\ \ref{fig:tpfhz} is shown here 
				as a function of time. The transformation from redshift to time is cosmology dependent. 
				To create this figure we have used best-fit values for the DDE parameters, $w_0=-1.0$ 
				and $w_a=-0.4$ \citep{Davis2007}. The y-axis is linear (c.f.\ the logarithmic axis in Fig.\ \ref{fig:tpfhz}) 
				and the family of curves have been re-normalized to highlight the sources of uncertainty 
				important for this analysis: uncertainty in the width of the function, and in the 
				location of its peak.
				The observer formation rate (OFR) is calculated by shifting the planet formation 
				rate by some amount $\Delta t_{obs}$ ($=4\ \textrm{Gyrs}$) to allow the planet to 
				cool, and the possible emergence of observers. These distributions are closed by 
				extrapolating exponentially in $t$.}
               \label{fig:tpfht}
       \end{center}
\end{figure}

The conversion from redshift to time depends on the 
particular cosmology, through the Friedmann equation,
	\begin{eqnarray}
		\left( \frac{da}{dt} \right)^2 & = & H(a)^2 a^2 \\
			& = & H_0^2 \bigg[ \Omega_{r 0}a^{-2} + \Omega_{m 0}a^{-1} + {} \biggr. \nonumber \\
			& & \biggl. \Omega_{de 0}\ \exp [3 w_a (a-1)]\ a^{-3 w_0 -3 w_a -1} \biggr]. \nonumber
	\end{eqnarray}
In Fig.\ \ref{fig:tpfht} we plot the PFR from Fig.\ \ref{fig:tpfhz} as a function of time assuming 
the best fit parameterized DDE cosmology.

\subsection{... then First Observers} 

After a star has formed, some non-trivial amount of time $\Delta t_{obs}$ will pass before observers,
if they arise at all, arise on an orbiting rocky planet. This time allows planets to form and cool and, 
possibly, biogenesis and the emergence observers. $\Delta t_{obs}$ is constrained to be shorter than the 
life of the host star. 
If we consider that our $\Delta t_{obs}$ has been drawn from a probability distribution $P_{\Delta t_{obs}}(t)$.
The observer formation rate (OFR) would then be given by the convolution
	\begin{equation} \label{eq:OFR}
		OFR(t) = const \times \int_{0}^{\infty} PFR(\tau) P_{\Delta t_{obs}}(t - \tau) d\tau.
	\end{equation}
	
In practice we know very little about $P_{\Delta t_{obs}}(t)$. It must be very nearly zero below
about $\Delta t_{obs} \sim 0.5\ \textrm{Gyrs}$ - this is the amount of time it takes for terrestrial 
planets to cool and the bombardment rate to slow down. Also, it is expected to be near zero above the 
lifetimes of sun-like stars (much above $\sim 10\ \textrm{Gyrs}$). 
If we assume 
that our $\Delta t_{obs}$ is typical, then $P_{\Delta t_{obs}}(t)$ has
significant weight around 
$\Delta t_{obs}=4\ \textrm{Gyrs}$ - the amount of time it has taken for us to evolve here on Earth. 

A fiducial choice, where \emph{all} observers emerge $4\ \textrm{Gyrs}$ after the formation 
of their host planet, is $P_{\Delta t_{obs}}(t)=\delta(t-4\ \textrm{Gyrs})$. This choice results in 
an OFR whose shape is the same as the PFR, but is shifted $4\ \textrm{Gyrs}$ into the future,
	\begin{equation}
			OFR(t) = const \times PFR(t - 4\ \textrm{Gyrs})
	\end{equation}
(see the lower panel of Fig.\ \ref{fig:tpfht}).	
Even for non-standard $w_0$ and $w_a$ values, this fiducial OFR aligns closely with the $r(t)$ peak 
and the effect of a wider $P_{\Delta t_{obs}}$ is generally to increase the severity of the coincidence 
problem by spreading observers outside the $r(t)$ peak. Hence using our fiducial $P_{\Delta t_{obs}}$ 
(which is the narrowest possibility) will lead to conclusions which are conservative in that they 
underestimate the severity of the cosmic coincidence. If another choice for $P_{\Delta t_{obs}}$ could 
be justified, the cosmic coincidence would be more severe than estimated here. We will discuss this 
choice in Section \ref{conclusion}.

The OFR is then extrapolated into the future using a decaying exponential with respect to $t$ (the 
dashed segment in the lower panel of Fig.\ \ref{fig:tpfht}). The observed SFH is consistent with 
a decaying exponential. 
We have tested other choices (linear \& 
polynomial decay) and our results do not depend strongly on the shape of the extrapolating function 
used. 

The temporal distribution of observers $P_t(t)$ is proportional to the observer formation rate,
	\begin{equation}
		P_t(t) = const \times OFR(t).
	\end{equation}

This observer distribution is similar to the one used by \citet{Garriga1999} to treat the coincidence 
problem in a multiverse scenario. By comparison, our $OFR(t)$ distribution starts later because we have 
considered the time required for the build up of metallicity, and because we have included an evolution 
stage of $4\ \textrm{Gyrs}$. Our distribution also decays more quickly than theirs does.
Some of our cosmologies suffer big-rip singularities in the future. In these cases we
truncate $P_t(t)$ at the big-rip.


\section{Analysis and Results: Does fitting contemporary constraints necessarily solve the cosmic coincidence?} \label{results}

\begin{figure}[!bt]
       \begin{center}
               \includegraphics[scale=0.52]{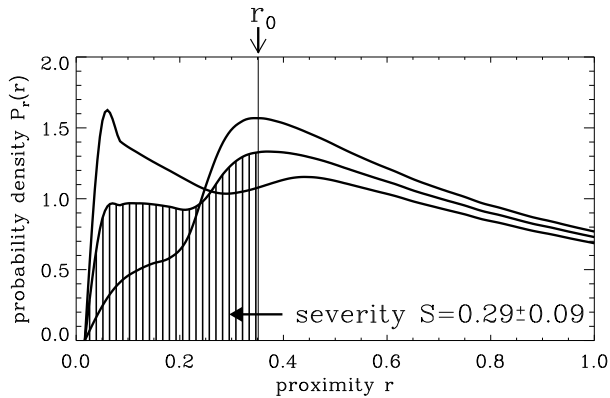}
               \caption{The predicted distribution of observations of $r$ is plotted for 
               the parameterized DDE model which best-fits cosmological observations: $w_0=-1.0$ 
               	and $w_a=-0.4$. The proximity parameter we observe 
				$r_0 =\frac{\rho_{m 0}}{\rho_{de 0}} \approx 0.35$ is typical in this cosmology since only
				$29 \%$ of observers (vertical striped area) observe $r<0.35$. The upper and lower 
				limits on this value resulting from 
				uncertainties in the SFR are $38 \%$ and $20 \%$ respectively. Thus the severity 
				of the cosmic coincidence in this model is $S=0.29\pm0.09$. This model does not suffer 
				a coincidence problem.}
               \label{fig:pofr}
       \end{center}
\end{figure}

For a given model the proximity parameter observed by a 
typical observer is described by a probability distribution $P_r(r)$ calculated as
	\begin{equation}
		P_r(r) = \sum \frac{dt}{dr} P_t(t(r)).
	\end{equation}
The summation is
over contributions from all solutions of $t(r)$ (typically,
any given value of $r$ occurs at multiple times during the lifetime of the Universe).
In Fig.\ \ref{fig:pofr} we plot $P_r(r)$ for the $w_0=-1.0$, $w_a=-0.4$ cosmology. In this case, 
observers are distributed over a wide range of $r$ values, with $71\%$
seeing $r>r_0$, and $29\%$ seeing $r<r_0$.

We define the severity $S$ of the cosmic coincidence problem as the 
probability that a randomly selected observer measures a proximity parameter 
$r$ lower than we do:
	\begin{equation}
		S = P(r < r_0) = 1 - P(r > r_0) = \int_{r=0}^{r_0} P_r(r) dr.
	\end{equation}
For the $w_0=-1.0$, $w_a=-0.4$ cosmology of Figs.\ $\ref{fig:tpfht}$ and $\ref{fig:pofr}$, the 
severity is $S=0.29 \pm 0.09$. This model does not suffer a coincidence problem since $29\%$
of observers would see $r$ lower than we do.
If the severity of the cosmic coincidence would be near $0.95$ ($0.997$) in a particular model, then 
that model would suffer a $2 \sigma$ ($3 \sigma$) coincidence problem and the value of $r$ we observe 
really would be unexpectedly high. 

We calculated the severities $S$ for cosmologies spanning a large region 
of the $w_0-w_a$ plane and show our results in Fig.\ \ref{fig:severities} using contours of
equal $S$. The severity of the coincidence problem is low (e.g.\ $S \lsim 0.7$) for most combinations
of $w_0$ and $w_a$ shown. There is a coincidence problem, where the severity is high ($S \gsim 0.8$), 
in two regions of this parameter space. These are indicated in Fig.\ \ref{fig:severities}. 

Some features in Fig.\ \ref{fig:severities} are worth noting: 
\begin{itemize}
\item Dominating the left of the plot, the severity of the coincidence increases
towards the bottom left-hand corner. This is because as $w_0$ and $w_a$ become more 
negative, the $r$ peak becomes narrower, and is observed by fewer observers. 
\item There is a strong vertical dipole of coincidence severity centered at $(w_0=0, w_a=0)$. 
For $(w_0 \approx 0, w_a > 0)$ there is a large coincidence problem because in such 
models we would be currently witnessing the very closest approach between DE and 
matter, with $\rho_{de} \gg \rhom$ for all earlier and later times (see Fig.\ \ref{fig:severityexamples}c). 
For $(w_0 \approx 0, w_a < 0)$ there is an anti-coincidence problem because in those models 
we would be currently witnessing the DDE's furthest excursion from the matter
density, with $\rho_{de}$ and $\rhom$ in closer proximity for all relevant earlier 
and later times, i.e., all times when $P_t(t)$ is non-negligible.
\item There is a discontinuity in the contours running along $w_a = 0$ for 
phantom models ($w_0 < -1$). The distribution $P_t(t)$ is truncated by 
big-rip singularities in strongly phantom models (provided they remain phantom; $w_a > 0$). 
This truncation of late-time observers means that early observers who witness large 
values of $r$ represent a greater fraction of the total population. 
\end{itemize}
 
To illustrate these features, Fig.\ \ref{fig:severityexamples} shows the density histories 
and observer distributions for four specific examples selected from the $w_0-w_a$ plane of 
Fig.\ \ref{fig:severities}.

We find that \emph{all} observationally allowed combinations of $w_0$ and $w_a$ result in low 
severities ($S < 0.4$), i.e., there are large ($> 60\%$) probabilities of observing the 
matter and vacuum density to be at least as close to each other as we observe them to be.

It is not suggested that for arbitrary models $P_{obs}$ should be large
when and only when $\rho_{de} \sim \rho_{m}$. Indeed, it is easy to imagine 
$\rho_{de} \sim \rho_{m}$ when there are not observers. Moreover, in some non-standard
cosmological models it is possible to contrive $\rho_{de} \ll \rho_{m}$ during times 
when observers do exist. What our results suggest is that, for models parameterized with $w_0$
and $w_a$ satisfying current constraints, most observers ($> 60\%$) will see $\rho_{de}$
and $\rho_m$ nearly equal ($r > 0.35$).


\section{Discussion} \label{conclusion}

\begin{figure*}[p!]
       \begin{center}
               \includegraphics[scale=1.0]{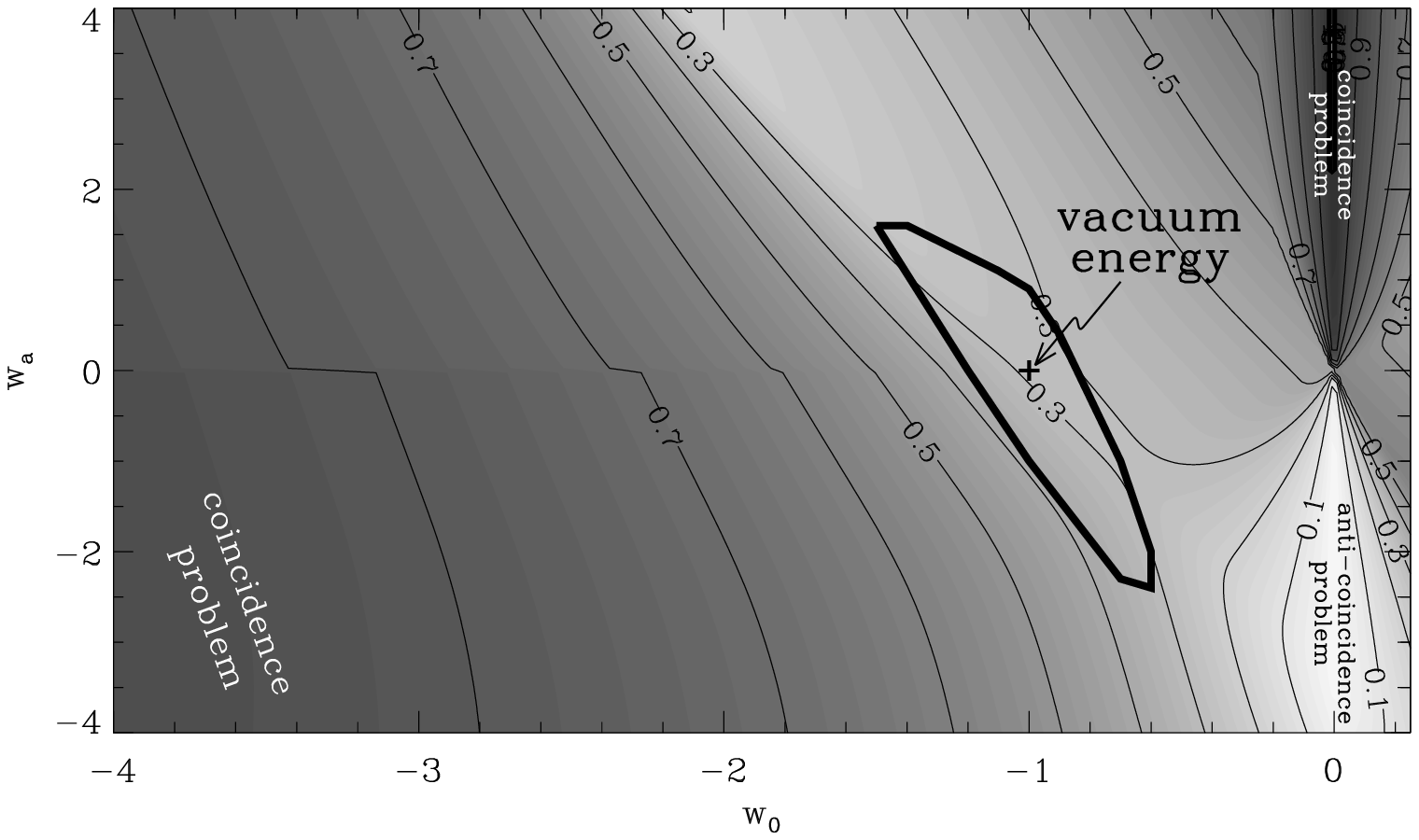}
               \caption{Here we plot contours of equal severity $S$ in $w_0-w_a$ parameter 
               space. $S$ is the fraction of observers who see $r < r_0$. If $S$ is large, 
               a large percentage of observers should see $r$ lower than we do - those models suffer 
               coincidence problems.
               The thick black contour represents the observational constraints on $w_0$ 
               and $w_a$ from \citet{Davis2007} ($2 \sigma$ confidence and marginalized over other uncertainties). 
               In \citet{Lineweaver2007} we showed that the severity of the coincidence problem is 
               low for \lcdm\ (indicated by the ``\textbf{+}''). 
               Values of $w_0$ and $w_a$ that result in a mild coincidence problem (e.g.\ $S \gsim 0.7$) 
               are already strongly excluded by observations. This leads to our main 
               result: none of the models in the observationally allowed regime 
               suffer a cosmic coincidence problem when our estimate of the temporal 
               distribution of observers $P_{obs}(t)$ is used as a selection function.
               }
               \label{fig:severities}
       \end{center}
\end{figure*}

\begin{figure*}[p!]
       \begin{center}
               \includegraphics[scale=1.0]{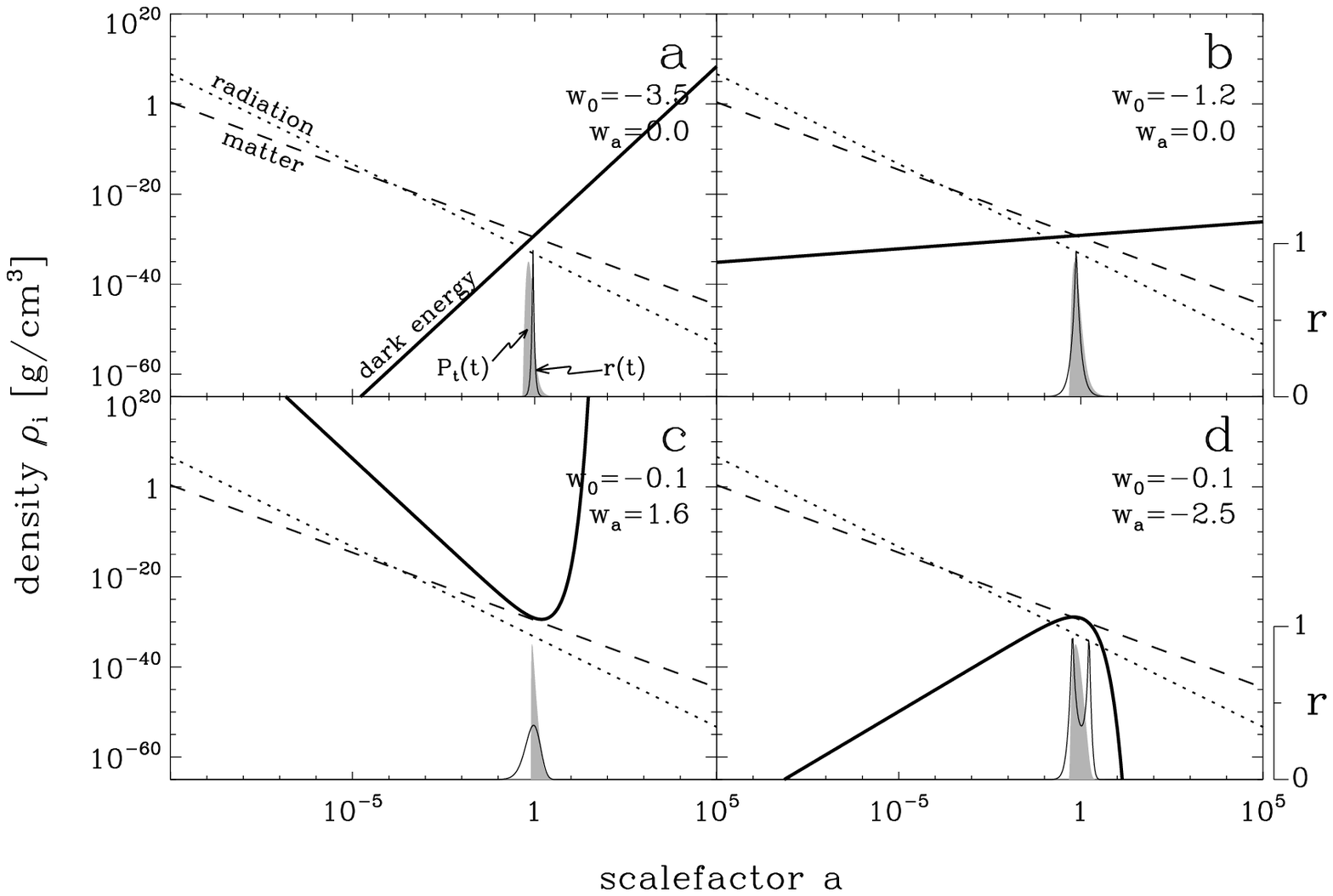}
               \caption{History of the energy densities in radiation (dotted line), matter (dashed line)
               and dark energy (thick black line) for four parameterized DE models from Fig.\ 
               \ref{fig:severities}. The proximity parameter $r$ (thin black line) and the temporal
               distribution of observers $P_t(t)$ (grey shade) are also given.
               {\bf Panel a} shows a phantom model with a constant equation of state
               $w=-3.5$. In this model the phantom density increases quickly and the $r(t)$ peak is 
               narrow. As a result, a large fraction of observers live while the matter and dark 
               energy densities are vastly different ($r \approx 0$) and there is a mild coincidence
               problem ($S \approx 0.8$). This might be used to rule-out the model shown in 
               Panel a, except that it is already strongly excluded by direct cosmological observations (refer 
               to Fig.\ \ref{fig:severities}). 
               {\bf Panel b} shows a phantom model which lies within the 
               observationally allowed $2\sigma$ region. There is no coincidence problem in 
               this model ($S \approx 0.4$). 
               {\bf Panel c} shows a model in which there is a 
               coincidence problem ($S \approx 0.95$). This models lies within the cluster of
               contours in the upper right-hand corner of Fig.\ \ref{fig:severities}. In this model
               the dark energy dominates the past \emph{and} future energy budget. Again however, the 
               coincidence problem can tell us nothing new, as this model is already strongly excluded 
               by observations.
               {\bf Panel d} shows a model in which there is an anti-coincidence problem. This models lies 
               within the cluster of contours in the lower right-hand corner of Fig.\ \ref{fig:severities}. 
               In this model the dark energy and matter densities are more similar ($r$ is greater) in the 
               recent past and near future (although $r \rightarrow 0$ further into the past or future). 
               According to the observer distribution $P_t(t)$ most observers live near the current epoch, 
               during $r>0.35$, with just $7 \%$ living during $r<0.35$ ($S=0.07$) in this particular model. 
               One might argue that this model can be ruled out because our value of $r$ is anomalously 
               small. However, this model too is already strongly excluded by observations.
	           }
               \label{fig:severityexamples}
       \end{center}
\end{figure*}

Particle-theoretic approaches to the cosmic coincidence
problem have focussed on the generation of a constant or slowly 
varying density ratio $r$. However it has not been made clear precisely 
how slowly the ratio $r$ must evolve in order to solve the coincidence
problem. In other words, the question ``What DDE dynamics are 
required to solve the coincidence problem?'' has not been addressed. 

In the present work we adopt the principle of mediocrity: that we 
should be typical observers, to try to answer this question.  
We estimate the temporal distribution of observers and devise a
scheme for quantifying how unlikely the observation $r \ge 0.35$ is for
an arbitrary DDE model. This scheme is applied to $w_0-w_a$ 
parametric models, and we identify regions of the $w_0-w_a$ parameter 
space in which the coincidence problem is most severe, however these are 
already strongly excluded by observations (see Fig.\ \ref{fig:severities}). 

Thus the main result of our analysis is that any realistic DDE model 
which can be parameterized as $w=w_0+w_a(1-a)$ over a few e-folds, 
has $\rho_{de} \sim \rho_m$ for a significant fraction of 
observers. 

Central to our approach is the temporal distribution of observers as
estimated using the distribution of terrestrial planets. Such 
observer selection effects are operating. Thus, while they may be
difficult to quantify, they need to be considered whenever the cosmic 
coincidence is used to motivate new physics. These anthropic 
considerations operate in conjunction with (not in place of) 
fundamental explanations of the dark energy.

%

Interacting quintessence models in which the proximity parameter asymptotes to a constant at 
late times 
\citep{Amendola2000,Zimdahl2001,Amendola2003,Franca2004,Guo2005,Olivares2005,Pavon2005,
Zhang2005,Franca2006,Amendola2006,Amendola2007,Olivares2007}
have been proposed as a solution to the coincidence problem. More recently, 
\citet{delCampo2005,delCampo2006} have argued for a broader class of interacting quintessence 
models that ``soften'' the coincidence problem by predicting a very slowly varying 
(though not constant) proximity parameter. 
Our analysis finds that $r$ need not asymptote to a constant, nor evolve particularly slowly, 
partially undermining the motivations for these interacting quintessence models.

\citet{Caldwell2003} and \citet{Scherrer2005} have proposed that the coincidence problem may 
be solved by phantom models in which there is a future big-rip singularity because such 
cosmologies spend a significant fraction of their lifetimes in $r \sim 1$
states. In our work $P_t(t)$ is terminated by big-rip singularities in ripping models. In 
non-ripping models, however, the distribution is effectively terminated by the declining star 
formation rate. Therefore the big-rip gives phantom models only a marginal advantage over other 
models. This marginal advantage manifests as the discontinuity along $w_a=0$ on the left side of 
Fig.\ \ref{fig:severities}.

A running cosmological constant $\Lambda(t)$ could arise from the renormalization group (RG) 
in quantum field theory \citep{Shapiro2000,Babic2002}. The running lambda term can mimic 
quintessence or phantom behaviour and transit smoothly between the two \citep{Sola2006}. RG models 
represent interesting alternatives to scalar-field models of dark energy. In some variants 
\citep{Grande2006b} additional fields are introduced to address the cosmic 
coincidence problem by predicting a slowly varying density ratio $r$. Our results demotivate such 
additions and favor simplistic RG models.

How strongly do these results depend on the assumed time it takes for observers to arise, 
$\Delta t_{obs}$? In \citet{Lineweaver2007}, where we performed an anlysis similar to the 
present one (but limited to $w=-1$), we demonstrated that the results were robust to any choice 
$\Delta t_{obs} \sim [0,11]\ \textrm{Gyrs}$. However, for $\Delta t_{obs} \gsim 12\ \textrm{Gyrs}$ 
that analysis resulted in an unavoidable coincidence problem because most observers would arise 
late (during DE domination) and would observe $\rho_{de} \gg \rho_{m}$. The validity of the 
results of the present analysis are similarly limited.

We could improve our analysis, in the sense of getting tighter coincidence constraints 
(larger severities), if we used a less conservative $P_{\Delta t_{obs}}$. We used the
most conservative choice - a delta function - because the present understanding of the
time it takes to evolve into observers is too poorly developed to motivate any other form of 
$P_{\Delta t_{obs}}$.
Another possible improvement is the DE equation of state parameterization. We used the
current standard, $w=w_0+w_a(1-a)$, which may not parameterize some models well for
very small or very large values of $a$.

We conclude that DDE models need not be fitted with exact tracking or oscillatory 
behaviors specifically to solve the coincidence by generating long or repeated periods of 
$\rho_{de} \sim \rho_{m}$. 
Also, particular interactions guaranteeing $\rho_{de} \sim \rho_m$ for 
long periods are not well motivated.
Moreover phantom models have no significant 
advantage over other DDE models with respect to the coincidence problem discussed here.


\section*{ACKNOWLEDGMENTS}
CE acknowledges a UNSW School of Physics postgraduate fellowship. CE thanks the ANU's RSAA 
for its kind hospitality, where this research was carried out.

\clearpage


\appendix
\section{Numerical Values for Parameters of Models Illustrated in Fig.\ \ref{fig:quintessence_energy}} \label{paramvals}

\begin{table}[hbt!]
   \begin{center}
   \caption{Free parameters of the DDE models
   illustrated in Fig.\ \ref{fig:quintessence_energy}. These values were chosen such 
   that observational constraints are crudely satisfied. These are by no means the 
   only combinations fitting observations. These values are intended for the purposes of
   illustration in Fig.\ \ref{fig:quintessence_energy}. Units are Planck units.}
   \begin{tabular}{l l l}
      \hline
      Model                            & Parameter                & Value \\
      \hline
      \hline
      power law tracker quintessence   & $\alpha$                 & $2$ \\
                                       & $M$                      & $1.4 \times 10^{-124}$ \\
      exponential tracker quintessence & $M$                      & $1.3 \times 10^{-124}$ \\
      tracking oscillating energy      & $M$                      & $1.8 \times 10^{-126}$ \\
                                       & $\lambda$                & $4$ \\
                                       & $A$                      & $0.99$ \\
                                       & $\nu$                    & $2.7$ \\
      interacting quintessence         & $A$                      & $1.4 \times 10^{-119}$ \\
                                       & $B$                      & $9.7$ \\
                                       & $C$                      & $16$ \\
      Chaplygin gas                    & $\alpha$                 & $1$ \\
                                       & $A$                      & $2.8 \times 10^{-246}$ \\                                      
      \hline
   \end{tabular}
   \label{tab:paramvals}
   \end{center}
\end{table}


\bibliographystyle{apsrev}
\bibliography{coincidenceII}



\end{document}